\documentclass[a4paper,11pt]{article}
\pdfoutput=1 

\usepackage{jheppub} 

\usepackage[T1]{fontenc} 

\title{\boldmath Accurate prediction of the $\phi^{*}_{\eta}$ distribution for the forward Z boson production in pp collisions through NNLO+N$^{3}$LL}

\author[a,1]{K. Ocalan,\note{Corresponding author.}}

\affiliation[a]{Faculty of Aviation and Space Sciences, Necmettin Erbakan University,\\ 42090 Meram Konya, Turkey}

\emailAdd{kadir.ocalan@erbakan.edu.tr}
\emailAdd{kadir.ocalan@cern.ch}

\abstract{This paper presents a phenomenological study for the differential cross section of the forward Z boson production in leptonic decay channels as a function of the angular variable $\phi^{*}_{\eta}$ in proton-proton collisions. The $\phi^{*}_{\eta}$ distribution is predicted for the forward pseudorapidity region $2.0 < \eta_{l} < 4.5$ of the decay leptons at center-of-mass energies 8, 13,  and 14 TeV. Accurate prediction of the $\phi^{*}_{\eta}$ distribution is achieved by means of the state-of-the-art calculations including fixed-order perturbative QCD and large logarithmic corrections. The predicted distributions are obtained by employing the resummation either at next-to-next-to-leading logarithmic (NNLL) or next-to-NNLL (N$^{3}$LL) accuracy which is matched to the perturbative QCD calculation at next-to-next-to-leading order (NNLO) accuracy, that is at NNLO+NNLL and NNLO+N$^{3}$LL, respectively. The Z boson $\phi^{*}_{\eta}$ variable is experimentally preferable as it has been measured more precisely than the Z boson transverse momentum though both variables probe the same physics, thereby an accurate description of the $\phi^{*}_{\eta}$ variable is required by using theoretical predictions at both NNLO+NNLL and NNLO+N$^{3}$LL accuracies. The matched predictions are compared with the available 8 TeV and 13 TeV measurements by the LHCb experiment at the LHC and found to be in good agreement. The 14 TeV predicted distributions at both NNLO+NNLL and NNLO+N$^{3}$LL are also reported. In all the predicted results, the NNLO+N$^{3}$LL provides more improved accuracy for the reliable description of the $\phi^{*}_{\eta}$ distribution throughout its entire phase space region.}

\begin{document} 
\maketitle
\flushbottom

\section{Introduction}
\label{intro}
The weak vector bosons, the W and Z boson, are copiously produced with clean experimental signatures in their leptonic decay channels in high-energy proton-proton (pp) collisions at the CERN LHC. Their leptonic decay processes provide benchmark tests of the standard model (SM) and substantial inputs for constraining the parton distribution functions (PDFs) of the proton. These processes are also of importance for modeling several SM and beyond the SM processes by means of constituting a non-negligible background as well as improving calibration for the detector response. Experimentally the Z boson is identified via its decays into pairs of electrons and muons, represented by $Z \rightarrow e^{-}e^{+}$ and $Z \rightarrow \mu^{-}\mu^{+}$ for the dielectron and dimuon final states, respectively\footnote{\label{myfootnote1}In this paper, $Z/\gamma^{*} \rightarrow l^{-}l^{+}$ process, where $l$ is an electron or a muon, is referred to as the Z boson process. The terms electron and muon are used to refer to both matter and anti-matter species of the particles.}. The Z boson can have nonzero transverse momentum $p_{T}$ due to the initial-state radiation of quarks and gluons and the intrinsic $p_{T}$ of the initial-state partons inside the proton. Measurement of the Z boson $p_{T}$ distribution provides important inputs for the SM precision measurements such as the measurement of the W boson mass~\cite{Aaboud:2017svj} and the background prediction of beyond the SM searches such as in the monojet topology~\cite{Aaboud:2017phn}. Moreover, measurement of the Z boson rapidity $y$ distribution in pp collisions is correlated with the longitudinal momentum fractions $x$ carried by the two interacting partons and provides constraints on the proton PDFs.

The Z boson $p_{T}$ and $y$ distributions were previously measured in both dielectron and dimuon decay modes in p$\rm{\bar{p}}$ collisions at center-of-mass energies 1.8 TeV and 1.96 TeV by the CDF and D0 Collaborations at the Tevatron~\cite{Affolder:1999jh, Aaltonen:2012fi, Abbott:1999yd, Abazov:2007ac, Abazov:2010kn, Abazov:2010mk}. More recently, the distributions were also measured in pp collisions such as at 8 TeV and 13 TeV by the ATLAS, CMS, and LHCb Collaborations at the LHC~\cite{Aad:2015auj, Aad:2019wmn, Khachatryan:2015oaa, CMS:2014jea, Khachatryan:2016nbe, Sirunyan:2019bzr, Aaij:2015zlq, AbellanBeteta:2016ugk, Aaij:2016mgv}. However, the Z boson $p_{T}$ measurements are limited in precision by the experimental uncertainties in the $p_{T}$ measurements of the decay leptons. The angular variable $\phi^{*}_{\eta}$~\cite{Banfi:2010cf, Banfi:2012du} was introduced to overcome this issue as an alternative probe of the Z boson $p_{T}$ with the following expression 
\begin{equation}
\phi^{*}_{\eta} = tan\left(\frac{\pi-\Delta \phi}{2}\right) sin(\theta^{*}_{\eta}), \hspace{0.5cm}cos(\theta^{*}_{\eta})=tanh\left(\frac{\Delta \eta}{2}\right),
\end{equation}
where $\Delta \phi$ and $\Delta \eta$ are the differences in azimuthal angle and pseudorapidity between the two leptons, respectively. The angle $\theta^{*}_{\eta}$ corresponds to the scattering angle of the lepton pairs relative to the proton beam direction in the rest frame of the dilepton sytem. The variable $\phi^{*}_{\eta}$ probes the same physics as the Z boson $p_{T}$ in terms of the approximate correlation $\phi^{*}_{\eta} \sim p_{T}/m_{ll}$, where $m_{ll}$ is the invariant mass of the lepton pair. The range $\phi^{*}_{\eta}\leq 1$ corresponds to Z boson $p_{T}$ up to about 100 GeV for a dilepton invariant mass close to the Z boson mass. The $\phi^{*}_{\eta}$ depends only on the angular direction of the leptons and is therefore measured more precisely than $p_{T}$ of the decay leptons due to the excellent spatial resolution of the detector systems. The Z boson $\phi^{*}_{\eta}$ distribution was previously measured by the D0 Collaboration~\cite{Abazov:2010mk}, and also by the ATLAS~\cite{Aad:2012wfa, Aad:2015auj, Aad:2019wmn}, CMS~\cite{Sirunyan:2017igm, Sirunyan:2019bzr}, and LHCb~\cite{Aaij:2012mda, Aaij:2015gna, Aaij:2015zlq, Aaij:2015vua, Aaij:2016mgv} Collaborations at 7, 8, and 13 TeV.

The total and differential cross sections of the Z boson have been predicted theoretically at next-to-next-to-leading order (NNLO) accuracy in perturbative QCD~\cite{Melnikov:2006kv, Catani:2009sm}. NNLO calculations for the weak vector boson production in association with a jet are also available~\cite{Ridder:2015dxa, Boughezal:2015ded, Boughezal:2015dva}. Electroweak corrections are particularly important at high-$p_{T}$ region of the Z boson which are known at next-to-leading order (NLO) accuracy~\cite{Dittmaier:2014qza, Lindert:2017olm}. Nevertheless, the fixed-order perturbative QCD calculations are unreliable at low $p_{T}$, where large logarithmic corrections are needed to be considered to account for soft and collinear gluon radiation~\cite{Collins:1984kg}. Next-to-next-to-leading logarithmic (NNLL) resummation of the logarithmically divergent terms has been matched with the fixed-order predictions to obtain accurate predictions for the $p_{T}$ spectrum~\cite{Balazs:1995nz, Catani:2015vma}. Parton shower models~\cite{Sjostrand:2014zea, Gleisberg:2008ta, Bahr:2008pv} can be used with fixed-order calculations to achieve fully exclusive predictions~\cite{Nason:2004rx, Frixione:2002ik, Alioli:2010xd, Alwall:2014hca}. Resummation and nonperturbative effects can also be incorporated by employing the transverse momentum dependent (TMD) PDFs~\cite{Angeles-Martinez:2015sea}. 

Precision measurements require accurate predictions of various angular and kinematical variables that match reduced experimental uncertainties from the vast amount of data collected at the LHC. Theoretical description of fiducial cross sections and kinematic distributions has been improved significantly by the NNLO QCD calculations. However, the fixed-order perturbative QCD calculations do not reliably describe differential distributions of the variables in the kinematical regions dominated by the soft and collinear QCD radiation. The perturbative expansion of cross section is affected by large logarithms in phase space regions dominated by soft and collinear radiation, therefore resummation of logarithmically enhanced terms to all orders in the strong coupling constant $\alpha_{s}$ is required to obtain physical description of variables. The Z boson $\phi^{*}_{\eta}$ ($p_{T}$) has been predicted at NNLO accuracy in perturbative QCD~\cite{Gehrmann-DeRidder:2016jns, Gehrmann-DeRidder:2017mvr}. The most accurate description of the Z boson $p_{T}$ and $\phi^{*}_{\eta}$ spectra has been achieved by the next-to-NNLL (N$^{3}$LL) resummation matched to NNLO prediction for the central detector acceptance of the decay leptons $|\eta_{l}|<2.5$ up to 13 TeV LHC pp collision energy~\cite{Bizon:2018foh, Bizon:2019zgf}.  

In this paper, the differential cross section predictions for the Z boson process in its leptonic decay modes are presented as a function of the $\phi^{*}_{\eta}$ in pp collisions. The $\phi^{*}_{\eta}$ distribution is predicted at the state-of-the-art accuracies including either NNLL or N$^{3}$LL resummation matched to the fixed-order perturbative QCD calculation at NNLO, referring to NNLO+NNLL and NNLO+N$^{3}$LL, respectively. It has been already shown that the NNLO calculations fail to describe $\phi^{*}_{\eta}$ distribution from the data~\cite{Aaij:2012mda}, where this distribution like that of $p_{T}$ is substantially affected by multiple soft gluon emissions which are not sufficiently accounted for in the fixed-order calculations. This observation clearly justifies the need for the merged predictions of this paper at NNLO+NNLL and NNLO+N$^{3}$LL for a reasonable description of the $\phi^{*}_{\eta}$ distribution. The merged predictions are obtained for the forward pseudorapidity region $2.0 < \eta_{l} < 4.5$ of the decay leptons apart from the central detector acceptance of the decay leptons $|\eta_{l}|<2.5$ considered in Refs.~\cite{Bizon:2018foh, Bizon:2019zgf}. Particularly the Z boson production in the forward region is of importance by means of probing effects at very low- and high-$x$ values and providing substantial input for constraining global PDFs~\cite{Harland-Lang:2014zoa, Ball:2014uwa, Dulat:2015mca}. The NNLO+NNLL and NNLO+N$^{3}$LL predictions for $\phi^{*}_{\eta}$ distribution are reported at 8 TeV and 13 TeV and compared with the available LHCb data in the forward region. The 14 TeV predicted distributions are also reported through the NNLO+N$^{3}$LL accuracy in this paper. Finally, the merged predictions for the Z boson $p_{T}$ distribution are included in the Appendix.      

\section{Methodology}
\label{meth}
\subsection{Computational setup}
\label{comp}
The calculations of fully differential cross sections including all-order resummation matched to fixed-order predictions are performed by using the MATRIX+RADISH (v1.0.0) computational framework~\cite{Kallweit:2020gva}. The fixed-order calculations in QCD perturbation theory are evaluated through the MATRIX framework~\cite{Grazzini:2017mhc, Catani:2009sm} which implements the $q_{T}$-subtraction method~\cite{Catani:2007vq, Catani:2012qa} for the cancellation of infrared divergences in the calculations. These divergences are regulated by introducing a fixed cut-off value $r_{cut}=$ 0.0015 (0.15\%) for the residual dependence parameter $r=p_{T}/m$, defined by the $p_{T}$ distribution and invariant mass $m$ for a system of colorless particles. The resummation of large logarithmic contributions is achieved with the formalism of the RADISH program~\cite{Bizon:2017rah, Monni:2016ktx} which is interfaced to the MATRIX framework. The RADISH code enables high-accuracy resummation of the transverse observables including $\phi^{*}_{\eta}$ and $p_{T}$. Moreover, all tree-level and one-loop amplitudes are acquired by means of the OpenLoops tool~\cite{Cascioli:2011va, Denner:2016kdg} through an automated interface in the computations. The calculations of differential cross sections in pp collisions require inclusion of knowledge of the PDFs. The LHAPDF 6.2.0 framework~\cite{Buckley:2014ana} is exploited for the evaluation of PDFs from data files in the computations. The NNPDF3.1~\cite{Ball:2014uwa} PDF set at NNLO accuracy is used in the calculations which is based on $\alpha_{s}=$ 0.118. 

\subsection{Fiducial phase space}
\label{fid}
The differential cross sections of the Born level variables $\phi^{*}_{\eta}$ and $p_{T}$ are calculated by using realistic phase space requirements for the decay products of the Z boson. The phase space requirements are directly taken from the Refs.~\cite{Aaij:2012mda, Aaij:2015gna, Aaij:2015zlq, Aaij:2015vua, Aaij:2016mgv} which were consistently used to define the fiducial acceptance of the LHCb measurements of the Z boson $\phi^{*}_{\eta}$ distribution at different pp collision energies. Leptons are treated massless in the computational framework of this paper, therefore the predicted differential cross section results in the dielectron channel are the same as in the dimuon channel. The leptons (either electrons or muons) are required to have $p_{T}>$ 20 GeV and to lie in the forward pseudorapidity region of $2.0 < \eta_{l} < 4.5$. In addition, the dilepton invariant mass $m_{ll}$ is required to be in the range $60 < m_{ll}<120$ GeV. No selection requirements are imposed for the final state hadronic jets. This fiducial phase space definition is consistently used in the calculations of this study performed at 8, 13, and 14 TeV pp collision energies.      

\subsection{Theoretical uncertainties}
\label{unc}
Perturbative QCD cross section calculations acquire dependence on the renormalization $\mu_{R}$ and factorization $\mu_{F}$ scales and therefore the numerical results depend on the choice of these scales. The central value for the scales is fixed to the Z boson mass $\mu_{R}$=$\mu_{F}$=$m(Z)$=91.1876 GeV. In the resummation component of cross section calculations, the central value for the resummation scale $x_{Q}$ is set to $x_{Q}=m/2$, where again $m$ stands for invariant mass of a system of colorless final states. The theoretical uncertainties due to $\mu_{R}$ and $\mu_{F}$ scale choices, referring to missing higher-order contributions in the perturbative QCD calculations, are estimated by varying independently the $\mu_{R}$ and $\mu_{F}$ up and down by a factor of 2 around the central value. The seven-point variation method is considered, that is all possible combinations in the variations are considered while imposing the limit $0.5 \leq \mu_{R}/ \mu_{F} \leq 2.0$ and keeping $x_{Q}=m/2$. In addition, $x_{Q}$ is varied around its central value by a factor of 2 in either direction for the central $\mu_{R}$ and $\mu_{F}$ scales. The theoretical uncertainties are defined as the envelope of the resulting nine-point scale variation. Thereafter, the total theoretical uncertainties are symmetrized by taking the larger values from estimated up and down uncertainties in a conservative approach and then reported in the predicted results of the $\phi^{*}_{\eta}$ ($p_{T}$) distribution throughout the entire paper. The clear focus is to report predicted results with the total theoretical uncertainty stemming from the combined calculation of the resummation and perturbative QCD expansion and thus theoretical uncertainties due to the choices of PDF set and $\alpha_{s}$ value are not considered in this work.

\section{Phenomenological results at 8 TeV and 13 TeV}
\label{pheno}
The differential cross sections for the Z boson process are calculated as a function of the $\phi^{*}_{\eta}$ by using the state-of-the-art merged predictions at NNLO+NNLL and NNLO+N$^{3}$LL accuracies in the forward region of 8 and 13 TeV pp collisions. The predicted $\phi^{*}_{\eta}$ distributions are compared with the available LHCb data~\cite{Aaij:2015vua, Aaij:2016mgv} for either the dielectron or dimuon decay mode in the fiducial phase space as defined in Sec.~\ref{fid}. Theoretical uncertainties due to the variations in the scales $\mu_{R}$, $\mu_{F}$, and $x_{Q}$ as discussed in Sec.~\ref{unc} are included in the predicted distributions. The data uncertainties are included by summing all sources of experimental uncertainties, that are reported in the related LHCb measurements, in quadrature. The $\phi^{*}_{\eta}$ distribution is binned as \{(0.00--0.01), (0.01--0.02), (0.02--0.03), (0.03--0.05), (0.05--0.07), (0.07--0.10), (0.10--0.15), (0.15--0.20), (0.20--0.30), (0.30--0.40), (0.40--0.60), (0.60--0.80), (0.80--1.20), (1.20--2.00), (2.00--4.00)\} in line with the LHCb $\phi^{*}_{\eta}$ measurements. The predicted results are obtained by using the NNPDF3.1 NNLO PDF sets. No nonperturbative and electroweak corrections are included as they are clearly left beyond the scope of this paper. 

\begin{figure}[tbp]
\centering 
\includegraphics[width=.65\textwidth]{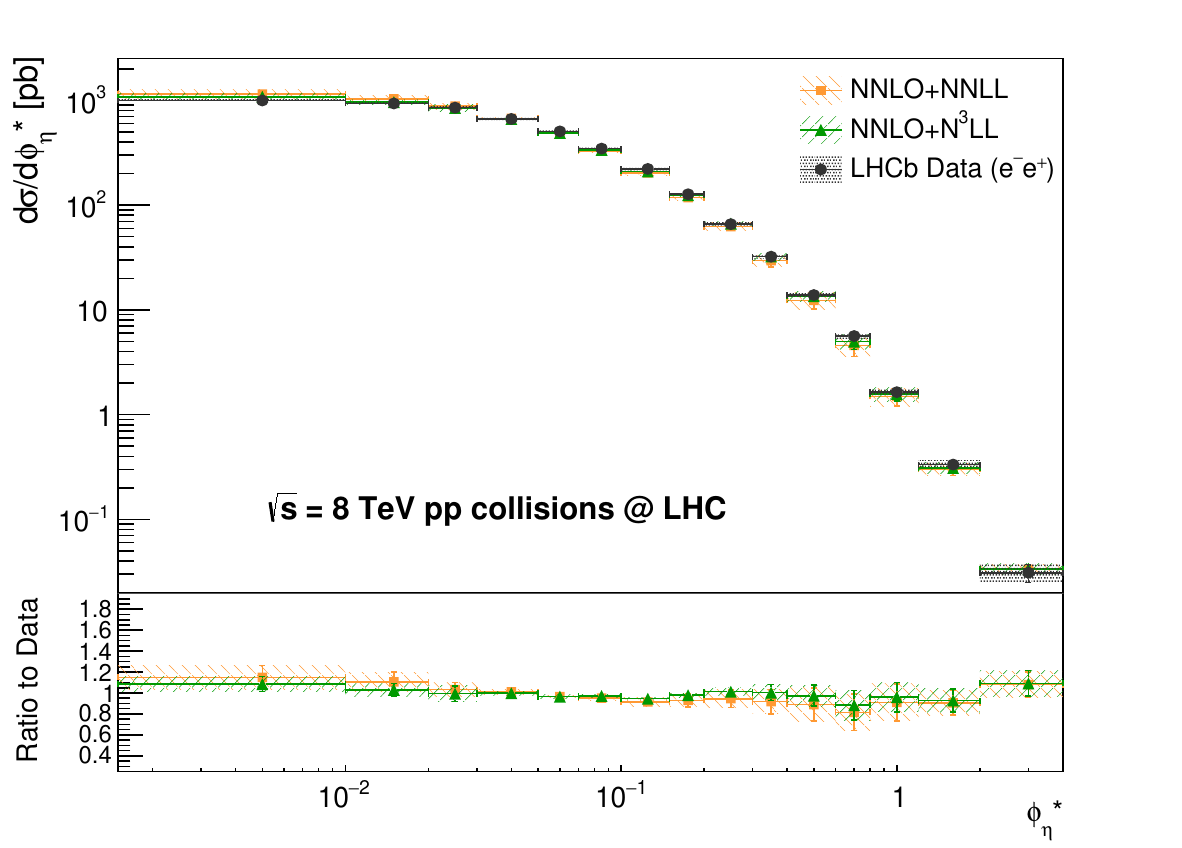}
\caption{\label{fig:1} The differential cross section distributions as a function of the $\phi^{*}_{\eta}$, $d\sigma/\phi^{*}_{\eta}$, for the Z boson process at 8 TeV. The predicted distributions at NNLO+NNLL and NNLO+N$^{3}$LL are compared with the LHCb data~\cite{Aaij:2015vua} in the dielectron decay mode. The predictions include theoretical uncertainties due to the scales as discussed in Sec.~\ref{unc}. The uncertainty that is included for the data is obtained by adding all sources of experimental uncertainties in quadrature. In the lower inset, the ratios of the predictions to the data for the $\phi^{*}_{\eta}$ are provided.}
\end{figure}

\begin{figure}[tbp]
\centering 
\includegraphics[width=.65\textwidth]{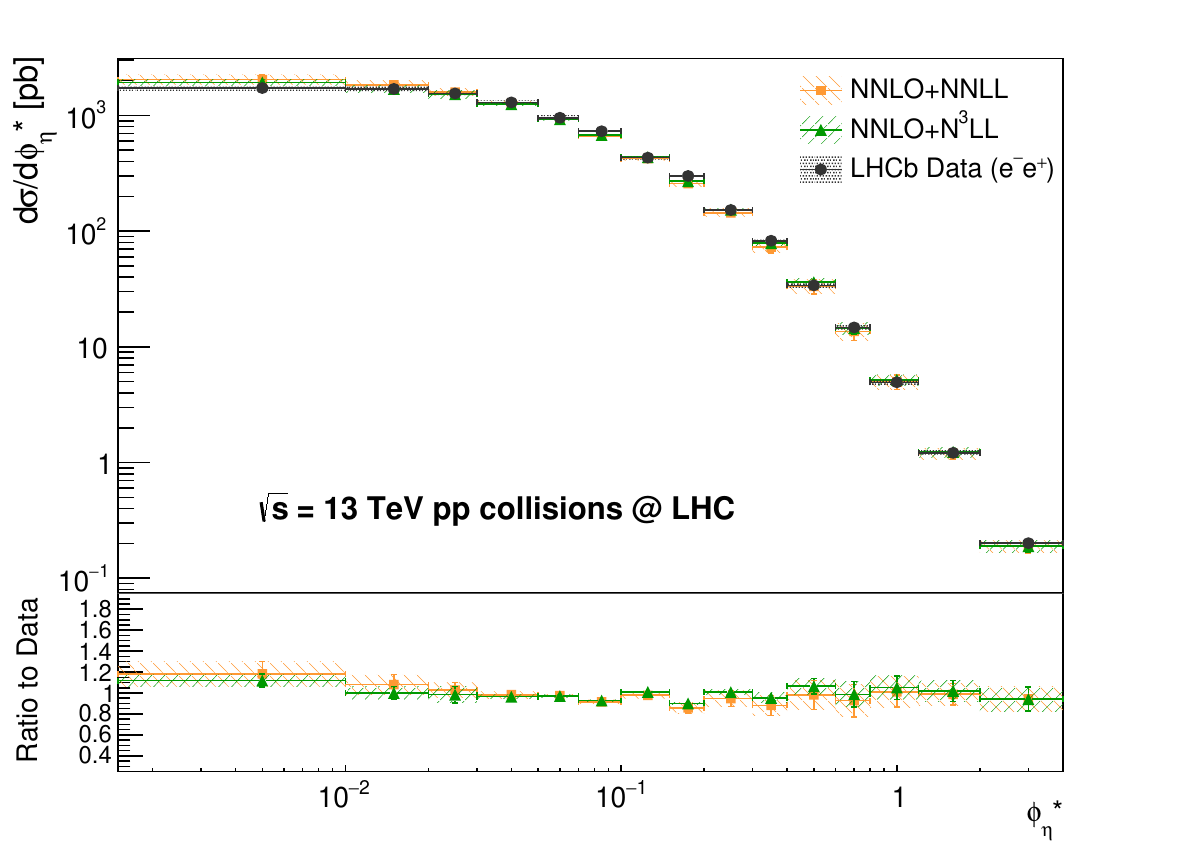}
\hfill
\includegraphics[width=.65\textwidth]{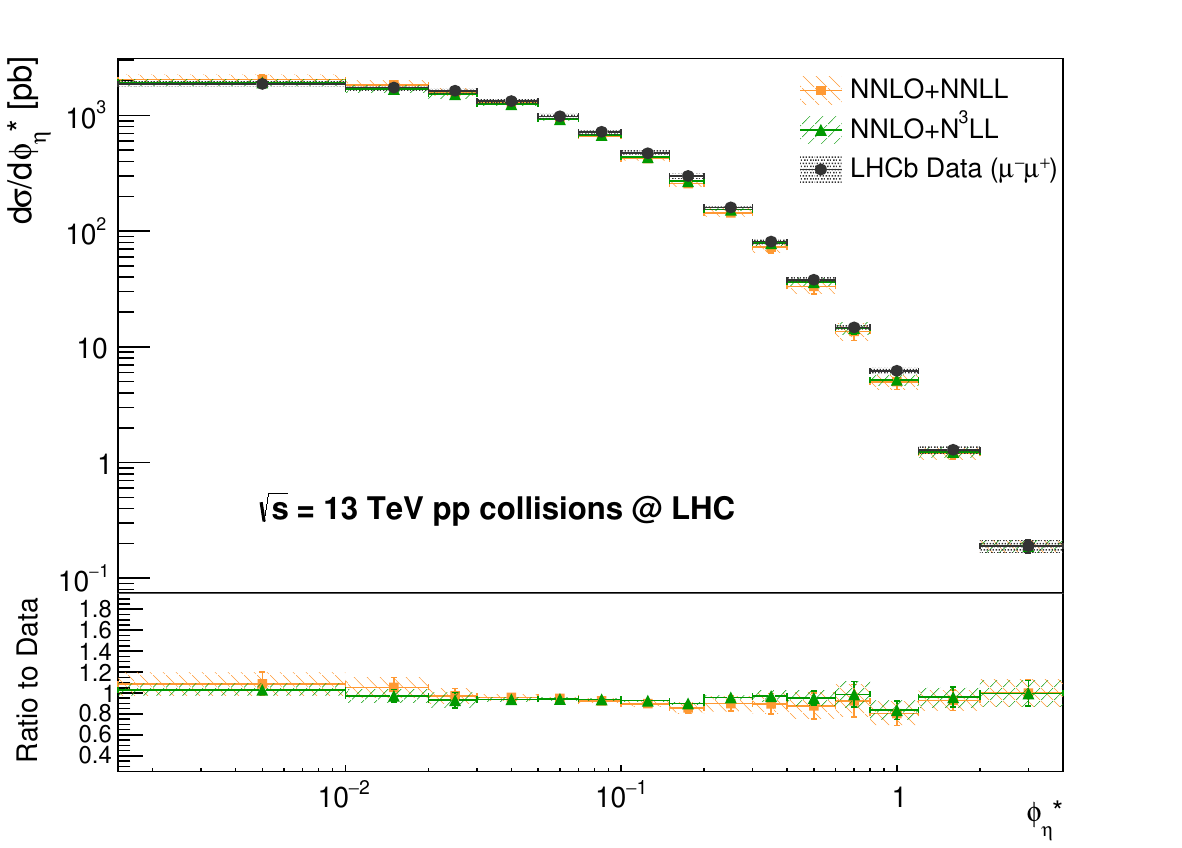}
\caption{\label{fig:2} The $d\sigma/\phi^{*}_{\eta}$ distributions for the Z boson process at 13 TeV. The predicted distributions at NNLO+NNLL and NNLO+N$^{3}$LL are compared with the LHCb data~\cite{Aaij:2016mgv} in both the dielectron (top) and dimuon (bottom) decay modes. The predictions include theoretical uncertainties due to the scales. The uncertainty that is included for the data is obtained by adding all sources of experimental uncertainties in quadrature. In the lower insets, the ratios of the predictions to the data for the $\phi^{*}_{\eta}$ are provided.}
\end{figure}

The 8 TeV $\phi^{*}_{\eta}$ distribution is predicted and compared with the LHCb data~\cite{Aaij:2015vua} in the dielectron decay mode as shown in figure~\ref{fig:1}. The NNLO+NNLL and NNLO+N$^{3}$LL predictions are both in good agreement with the data within uncertainties for the entire $\phi^{*}_{\eta}$ region. The NNLO+N$^{3}$LL provides more accurate description of the data with less than 4.0\% deviation. The predictions are able reproduce the data for the low-$\phi^{*}_{\eta}$ region ($\phi^{*}_{\eta}<$ 0.1), where the resummation components of the predictions also provide reliable results. The NNLO+NNLL tends to overestimate slightly the data in the low-$\phi^{*}_{\eta}$ region and underestimate slightly the data in the intermediate-$\phi^{*}_{\eta}$ region, while description of the data in these regions is improved more by the NNLO+N$^{3}$LL prediction. The 13 TeV $\phi^{*}_{\eta}$ distribution is also predicted and compared with the corresponding LHCb data~\cite{Aaij:2016mgv} in both the dielectron and dimuon decay modes as shown in figure~\ref{fig:2}. The merged predictions are able to reproduce data well in both the dielectron and dimuon decay modes within uncertainties for almost the entire $\phi^{*}_{\eta}$ ranges. In only a few $\phi^{*}_{\eta}$ bins, the level of disagreement with the data is up to 9\% in both the decay modes. The NNLO+N$^{3}$LL prediction provides better agreement with the data as anticipated. In the low-$\phi^{*}_{\eta}$ region, description of the data is improved with the inclusion of the resummation at N$^{3}$LL to the NNLO QCD prediction as compared to the NNLO+NNLL prediction, where it tends to overestimate more the dielectron and dimuon data.     

\begin{figure}[tbp]
\centering 
\includegraphics[width=.45\textwidth]{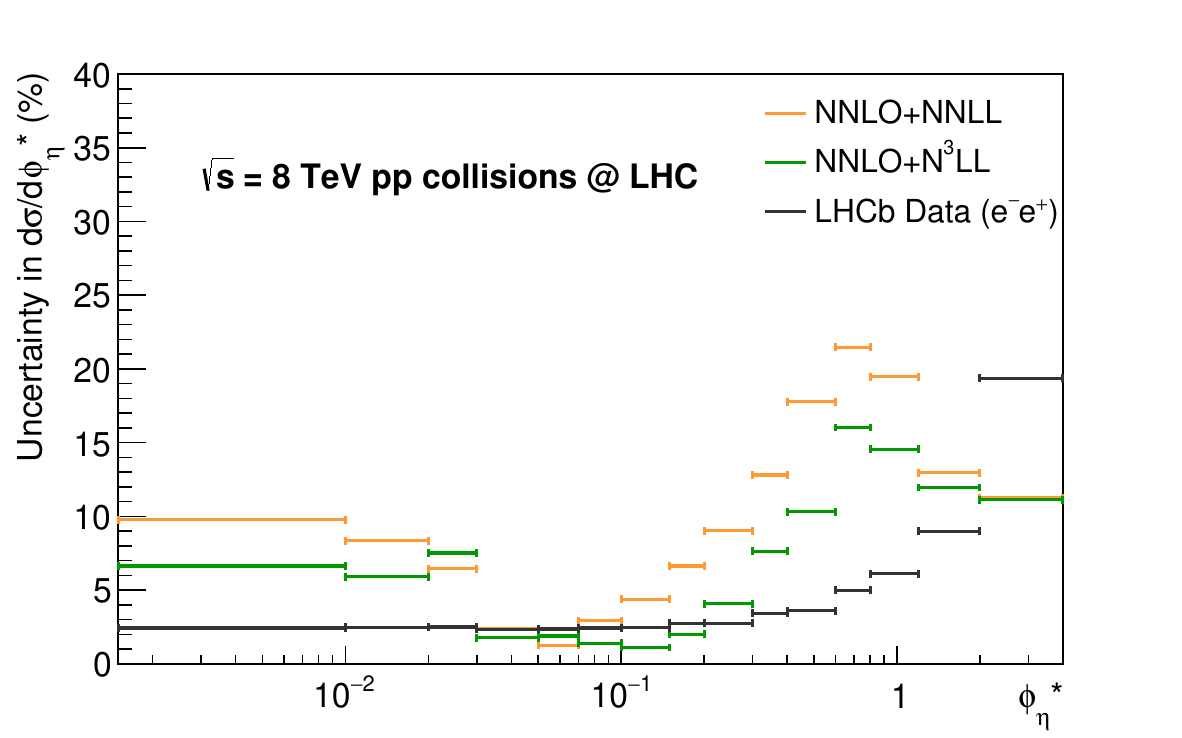}
\hfill
\includegraphics[width=.45\textwidth]{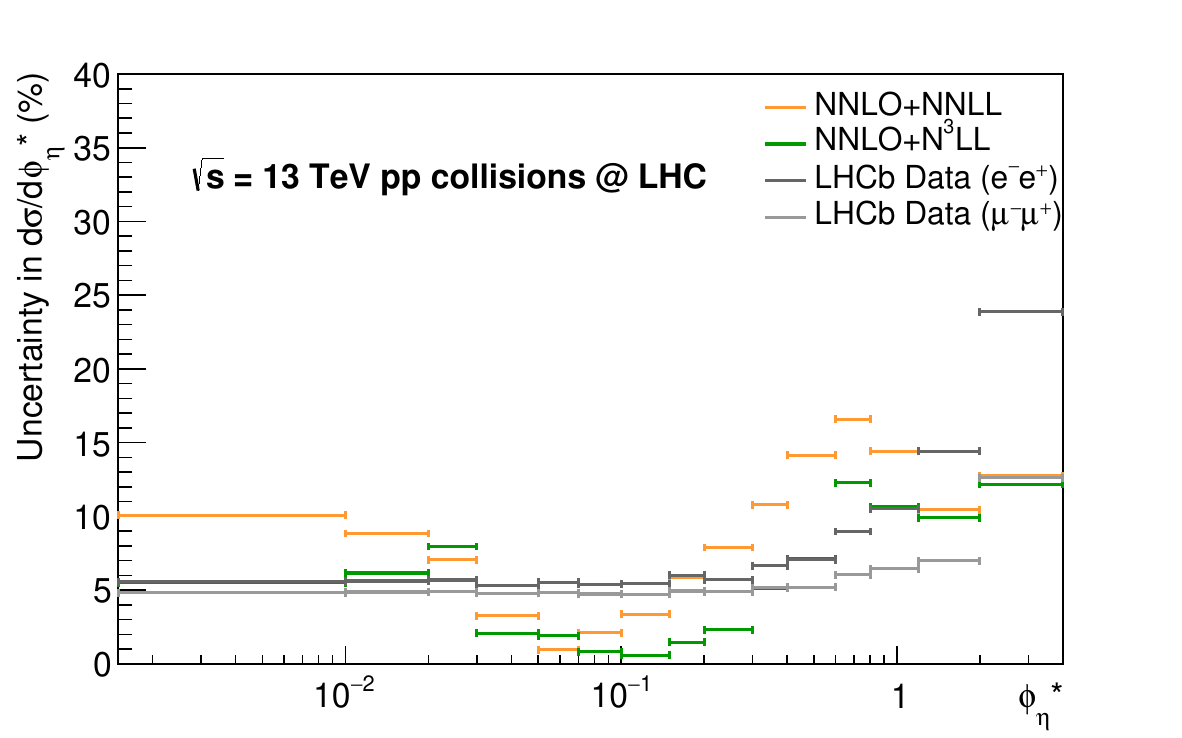}
\caption{\label{fig:3} Comparisons of the estimated theoretical uncertainties at NNLO+NNLL and NNLO+N$^{3}$LL with the total experimental uncertainties for the $d\sigma/\phi^{*}_{\eta}$ distributions in percent at 8 TeV (left) and 13 TeV (right). Theoretical uncertainties due to the scales are estimated by following the procedure as discussed in Sec.~\ref{unc}. Total experimental uncertainties are calculated by adding all sources of experimental uncertainties, that are reported in the LHCb measurements~\cite{Aaij:2015vua, Aaij:2016mgv}, in quadrature.}
\end{figure}

In both the 8 TeV and 13 TeV results, the precision achieved by the NNLO+N$^{3}$LL prediction is significantly higher than the NNLO+NNLL prediction as can be seen in figure~\ref{fig:3}. Despite theoretical uncertainties of the predictions are comparable in some $\phi^{*}_{\eta}$ bins, they are mainly reduced in the low-$\phi^{*}_{\eta}$ and intermediate- to high-$\phi^{*}_{\eta}$ bins in the N$^{3}$LL resummation. Theoretical uncertainties are reduced to a few percent in bins around $\phi^{*}_{\eta}=$ 0.1 and become maximum around $\phi^{*}_{\eta}=$ 1 towards the Z boson $p_{T}\approx$ 100 GeV. Moreover, the estimated theoretical uncertainties of the NNLO+N$^{3}$LL prediction are lower than the total experimental uncertainties in some $\phi^{*}_{\eta}$ ranges such as 0.03--0.20 and 2.0--4.00 at 8 TeV and 0.03--0.3 and 1.20--4.00 at 13 TeV.      

\section{Merged predictions at 14 TeV}
\label{merged}
\begin{figure}[tbp]
\centering 
\includegraphics[width=.65\textwidth]{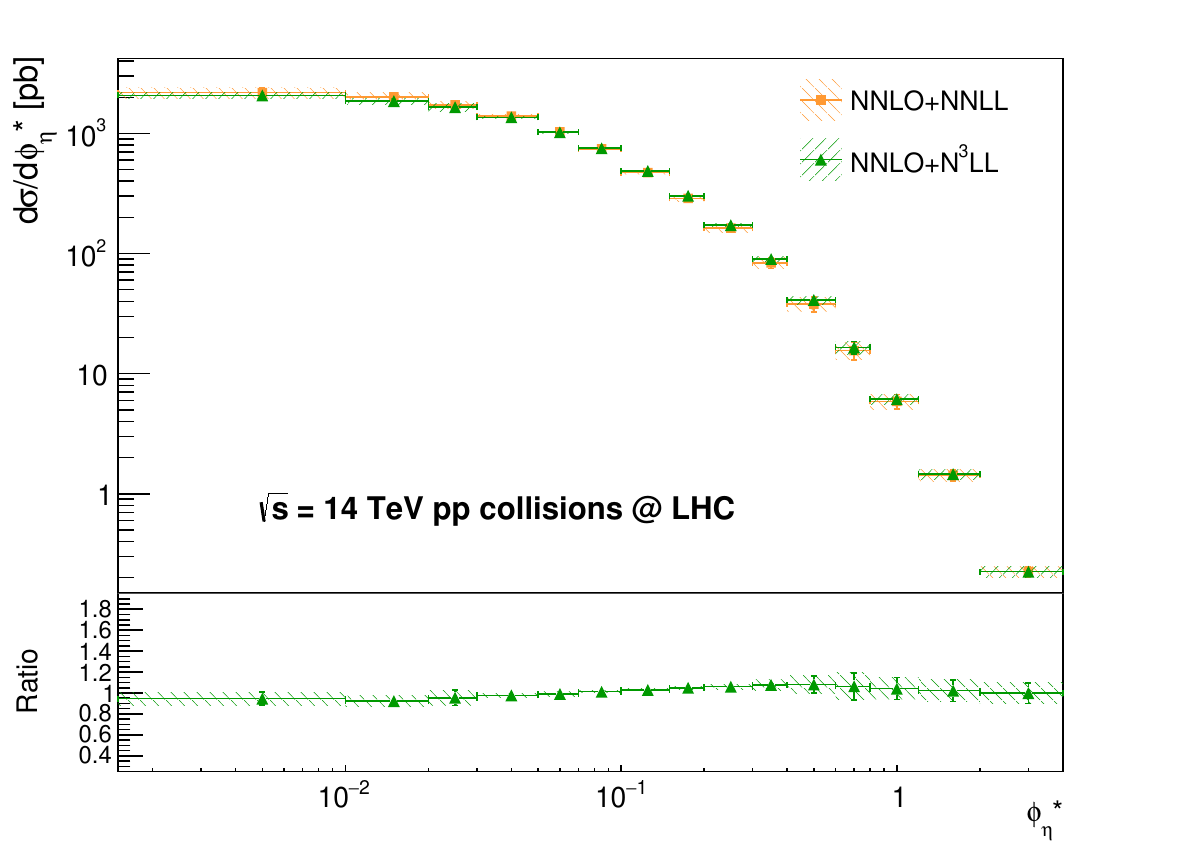}
\caption{\label{fig:4} The $d\sigma/\phi^{*}_{\eta}$ distributions for the Z boson process at 14 TeV which are predicted at NNLO+NNLL and NNLO+N$^{3}$LL. The predictions include theoretical uncertainties due to the scales. In the lower inset, the ratio of the predictions NNLO+N$^{3}$LL-to-NNLO+NNLL for the $\phi^{*}_{\eta}$ is provided.}
\end{figure}

\begin{figure}[tbp]
\centering 
\includegraphics[width=.55\textwidth]{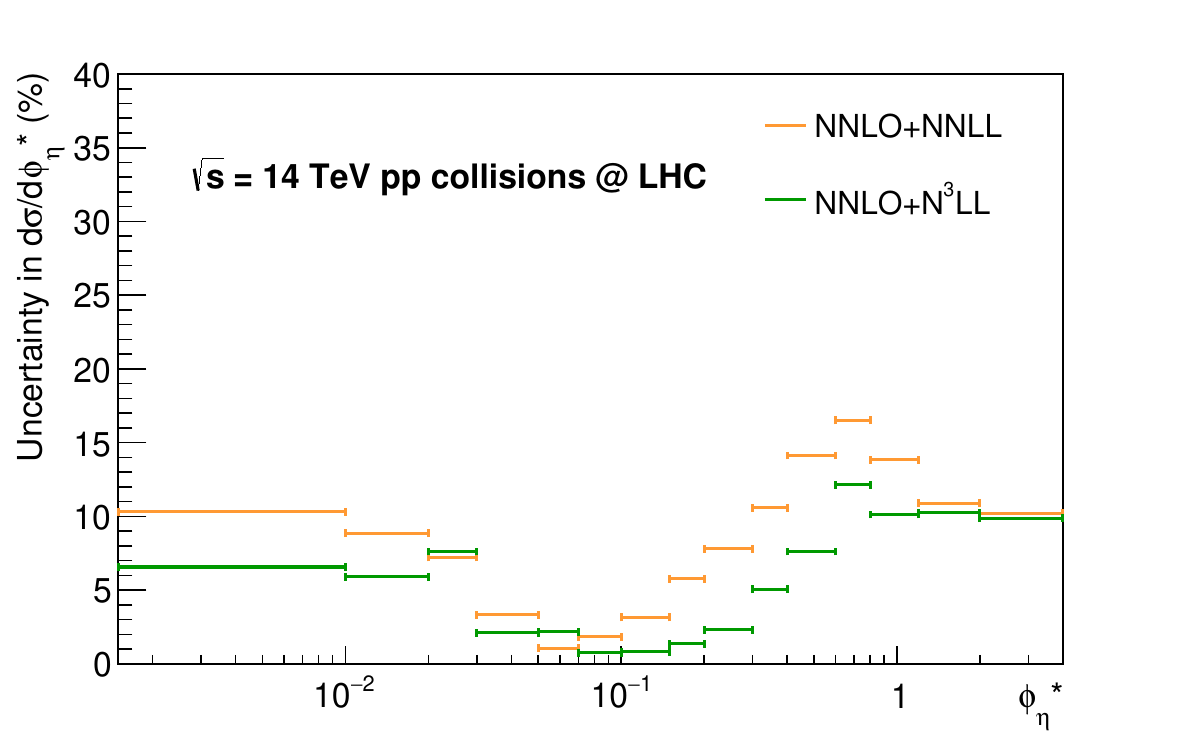}
\caption{\label{fig:5} Comparisons of the estimated theoretical uncertainties at NNLO+NNLL and NNLO+N$^{3}$LL for the $d\sigma/\phi^{*}_{\eta}$ distributions in percent at 14 TeV. Theoretical uncertainties due to the scales are estimated by following the procedure as discussed in Sec.~\ref{unc}.}
\end{figure}

The NNLO+NNLL and NNLO+N$^{3}$LL predictions at 8 TeV and 13 TeV are justified with the corresponding LHCb data for the $\phi^{*}_{\eta}$ in Sec.~\ref{pheno}. The 8 TeV and 13 TeV predictions through the resummation at N$^{3}$LL incorporated to the NNLO QCD prediction are found to describe the data accurately which further encourage the extension of the predictions to 14 TeV center-of-mass energy in the forward region of pp collisions. The 14 TeV $\phi^{*}_{\eta}$ distribution is also predicted based on the same methodology of the 8 TeV and 3 TeV predictions encompassing the computational setup, fiducial phase space definition for the Z boson process, binning choice, and procedure for the estimation of theoretical uncertainties. The $\phi^{*}_{\eta}$ distribution is compared between the NNLO+NNLL and NNLO+N$^{3}$LL predictions at 14 TeV as shown in figure.~\ref{fig:4}. The predictions are observed to be consistent with each other within the uncertainties for the entire $\phi^{*}_{\eta}$ region, where the NNLO+N$^{3}$LL prediction is slightly lower in the low-$\phi^{*}_{\eta}$ and higher in the intermediate-$\phi^{*}_{\eta}$ region in comparison to the NNLO+NNLL prediction which assures reliable description of the distribution with the inclusion of the resummed logarithmic terms at N$^{3}$LL. In addition, the theoretical uncertainties estimated for the predictions of the $\phi^{*}_{\eta}$ distribution are compared in figure.~\ref{fig:5}. The precision achieved by the NNLO+N$^{3}$LL is higher over the NNLO+NNLL for almost the entire $\phi^{*}_{\eta}$ region as one would also anticipate.        

To this end, the 14 TeV predicted results are compared to the 8 TeV and 13 TeV predictions in terms of the numerical values of the differential cross sections and the cumulative cross sections obtained from the differential distributions. The predicted numerical values of the differential cross sections at NNLO+N$^{3}$LL in bins of the $\phi^{*}_{\eta}$ are provided in table~\ref{tab:1}. The predicted values increase significantly in going from a lower energy to 14 TeV in each bin. The predicted cumulative cross section at 14 TeV is compared to the predictions at 8 TeV and 13 TeV in table~\ref{tab:2}. The NNLO+N$^{3}$LL predicts higher cumulative cross sections apart from the 14 TeV results.  

\begin{table}[tbp]
\centering
\begin{tabular}{|c|c|c|c|} 
\hline
Bins & 8 TeV & 13 TeV & 14 TeV  \\
\hline
0.00--0.01 & 1080.91$\pm$71.64 & 1930.19$\pm$106.09 & 2073.47$\pm$136.18 \\
0.01--0.02 & 960.87$\pm$56.64 & 1695.40$\pm$104.32 & 1858.40$\pm$109.62 \\
0.02--0.03 & 845.60$\pm$63.52& 1524.11$\pm$121.25 & 1655.40$\pm$126.06 \\ 
0.03--0.05 & 662.85$\pm$11.70& 1250.11$\pm$25.87 & 1362.97$\pm$28.76 \\
0.05--0.07 & 487.56$\pm$9.19 & 927.31$\pm$17.76 & 1022.98$\pm$22.34 \\
0.07--0.10 & 336.33$\pm$4.69 & 677.87$\pm$5.50 & 753.38$\pm$5.65 \\ 
0.10--0.15 & 209.43$\pm$2.31 & 436.53$\pm$2.54 & 486.65$\pm$3.92 \\
0.15--0.20 & 124.29$\pm$2.49 & 270.19$\pm$3.98 & 302.63$\pm$4.11 \\
0.20--0.30 & 66.63$\pm$2.70 & 153.63$\pm$3.55 & 172.40$\pm$3.97 \\ 
0.30--0.40 & 32.32$\pm$2.46 & 78.78$\pm$4.02 & 90.11$\pm$4.56 \\
0.40--0.60 & 13.50$\pm$1.39 & 36.19$\pm$2.58 & 41.06$\pm$3.13 \\
0.60--0.80 & 4.97$\pm$0.80 & 14.53$\pm$1.78 & 16.57$\pm$2.01 \\ 
0.80--1.20 & 1.57$\pm$0.23 & 5.20$\pm$0.55 & 6.13$\pm$0.62 \\
1.20--2.00 & 0.31$\pm$0.04 & 1.24$\pm$0.12 & 1.46$\pm$0.15 \\
2.00--4.00 & 0.034$\pm$0.004 & 0.190$\pm$0.023 & 0.225$\pm$0.022 \\ 
\hline
\end{tabular}
\caption{\label{tab:1} The predicted $d\sigma/\phi^{*}_{\eta}$ values at NNLO+N$^{3}$LL in the fiducial phase space of the Z boson process at 8 TeV, 13 TeV, and 14 TeV in bins of the $\phi^{*}_{\eta}$. Theoretical uncertainties due to the scales are included to the central results.}
\end{table}

\begin{table}[tbp]
\centering
\begin{tabular}{|c|c|c|c|} 
\hline
Accuracy & 8 TeV & 13 TeV & 14 TeV  \\
\hline
NNLO+NNLL  & 93.105$\pm$1.215 pb & 187.221$\pm$2.423 pb & 207.542$\pm$2.884 pb \\ 
\hline
NNLO+N$^{3}$LL & 93.193$\pm$1.197 pb & 187.549$\pm$2.424 pb & 207.502$\pm$2.847 pb \\
\hline
\end{tabular}
\caption{\label{tab:2} The predicted cumulative cross section values at both NNLO+NNLL and NNLO+N$^{3}$LL in the fiducial phase space of the Z boson process at 8 TeV, 13 TeV, and 14 TeV for the $\phi^{*}_{\eta}$ range 0.0--4.0. Theoretical uncertainties due to the scales are included to the central results.}
\end{table}

\section{Summary and conclusion}
\label{conc}
In this paper, the differential cross section predictions are presented as a function of the $\phi^{*}_{\eta}$ variable, which is related to the $p_{T}$ of the Z boson. The predictions are presented for the Z boson decaying leptonically in the forward region of pp collisions at center-of-mass energies 8, 13, and 14 TeV. The predictions for the $\phi^{*}_{\eta}$ distribution are reported by employing the fiducial phase space definition of the related LHCb measurements  which encompasses the forward detector region. The fiducial phase space definition requires the leptons (either electrons or muons) to have $p_{T}>$ 20 GeV, the dilepton invariant mass $m_{ll}$ to be in the window $60 < m_{ll}<120$ GeV, and the forward region to be $2.0 < \eta_{l} < 4.5$ in terms of the pseudorapidities of the decay leptons. The $\phi^{*}_{\eta}$ distribution is obtained from the state-of-the-art predictions of the next-to-next-to-leading order (NNLO) perturbative QCD calculations combined with the resummation of the logarithmically enhanced terms at either next-to-next-to-leading logarithm (NNLL) or next-to-NNLL (N$^{3}$LL), NNLO+NNLL and NNLO+N$^{3}$LL, respectively. Theoretical uncertainties due to the perturbation and resummation scales of the merged predictions NNLO+NNLL and NNLO+N$^{3}$LL are also estimated and reported for the $\phi^{*}_{\eta}$ distribution. 
      
The predicted $\phi^{*}_{\eta}$ distributions are compared with the available LHCb measurements~\cite{Aaij:2015vua, Aaij:2016mgv} either in the dielectron or dimuon decay modes at 8 TeV and 13 TeV. The NNLO+NNLL and NNLO+N$^{3}$LL merged predictions are both found to be in very good agreement with the data within the uncertainties throughout the entire $\phi^{*}_{\eta}$ region. The NNLO+N$^{3}$LL provides much better agreement with the data in terms of the accuracy achieved where the deviations are less than 4.0 (9.0)\% at 8 TeV (13 TeV) in comparison to the NNLO+NNLL. The data description is improved considerably with the inclusion of the resummation at N$^{3}$LL to the fixed-order NNLO QCD prediction over the NNLO+NNLL prediction, particularly for the low-$\phi^{*}_{\eta}$ region $\phi^{*}_{\eta}<$ 0.1. Overall, the precision achieved in the NNLO+N$^{3}$LL is significantly higher than the NNLO+NNLL for most the $\phi^{*}_{\eta}$ ranges. The 8 TeV and 13 TeV predictions provide reliable description of the data with high accuracy for the $\phi^{*}_{\eta}$ variable and justify the extension of these state-of-the-art predictions to the 14 TeV. The 14 TeV $\phi^{*}_{\eta}$ distribution is consistently predicted by the NNLO+NNLL and NNLO+N$^{3}$LL within the uncertainties for the entire $\phi^{*}_{\eta}$ region. The NNLO+N$^{3}$LL prediction exhibits slightly lower (higher) distribution in the low-$\phi^{*}_{\eta}$ (the intermediate-$\phi^{*}_{\eta}$) region in comparison to the NNLO+NNLL prediction, which assures more accurate description of the distribution with the inclusion of the resummed logarithmic terms at N$^{3}$LL. Finally, this paper showed that the merged predictions at NNLO+NNLL and NNLO+N$^{3}$LL are required to model reliably the $\phi^{*}_{\eta}$ distribution of the forward Z boson production in pp collisions. The merged predictions presented in this paper are recommended for theoretical comparisons of experimental data in future Z boson $\phi^{*}_{\eta}$ and $p_{T}$ measurements at the LHC.   

\appendix
\section{Merged predictions for the forward Z boson $p_{T}$ distribution}
\begin{figure}[tbp]
\centering 
\includegraphics[width=.55\textwidth]{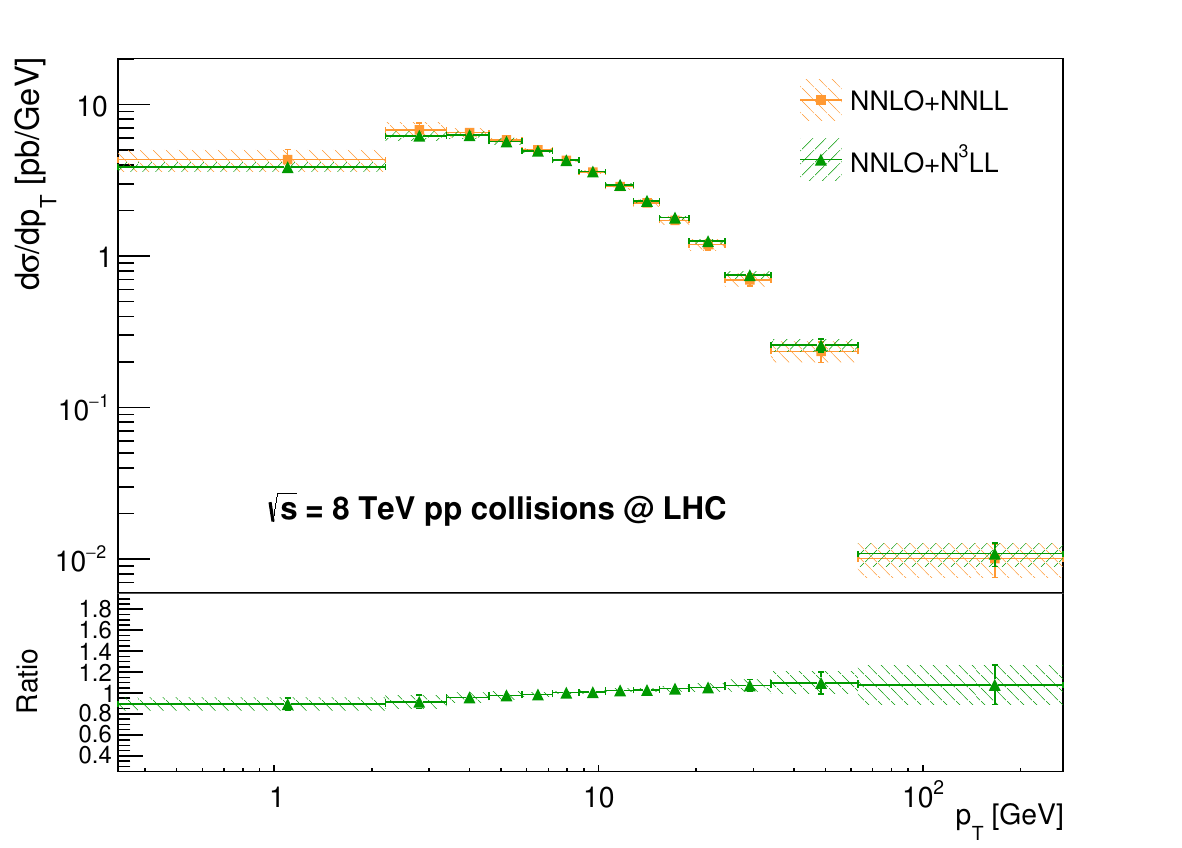}
\hfill
\includegraphics[width=.55\textwidth]{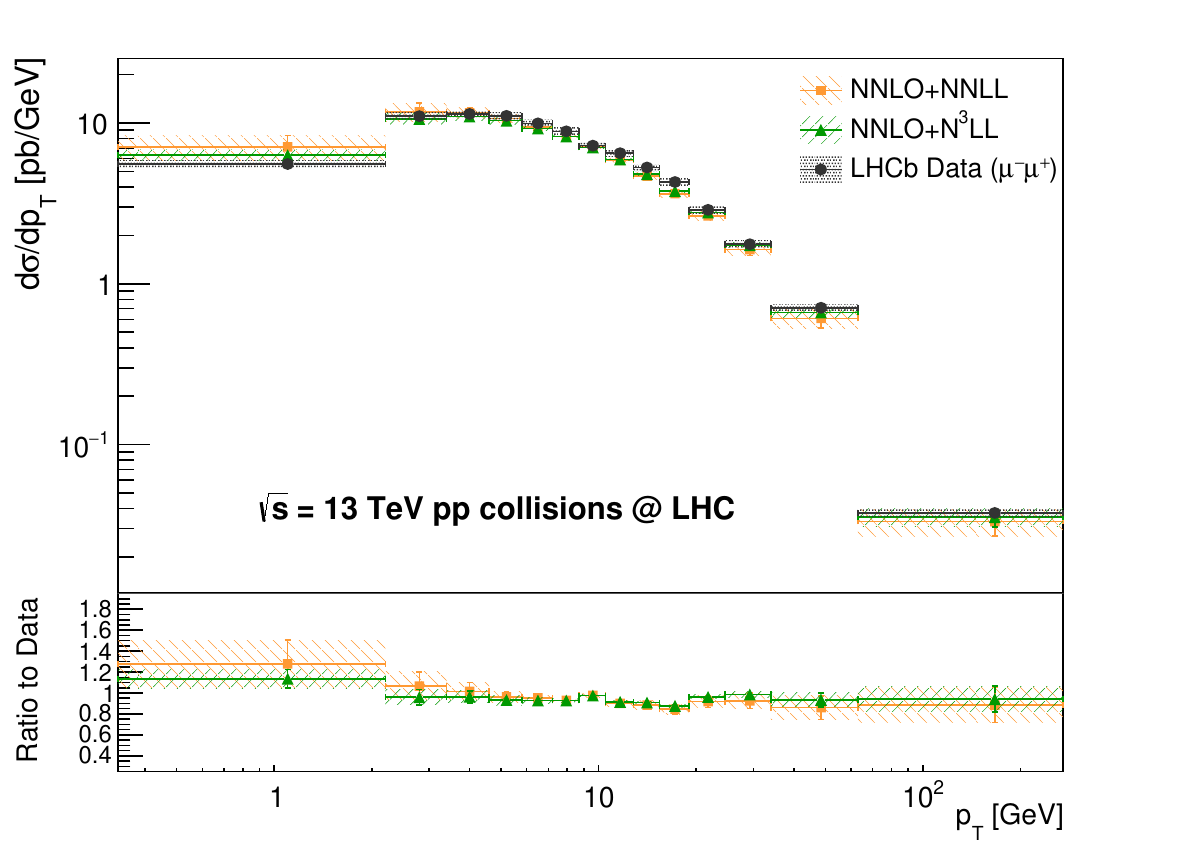}
\hfill
\includegraphics[width=.55\textwidth]{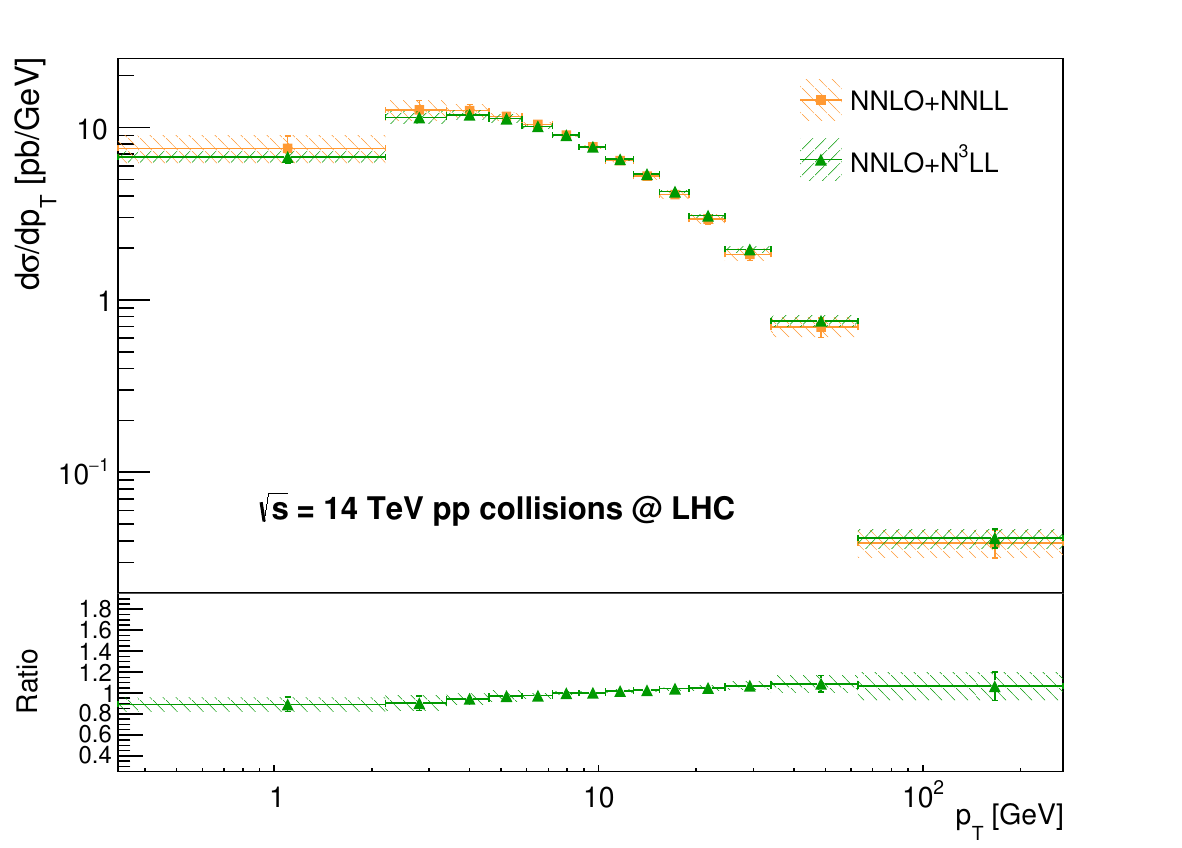}
\caption{\label{fig:6} The predicted $d\sigma/\phi^{*}_{\eta}$ distributions for the forward Z boson $p_{T}$ up to 270 GeV at 8 TeV (top), 13 TeV with the comparison of the LHCb dimuon data~\cite{Aaij:2016mgv} (middle), and 14 TeV (bottom). The predictions include theoretical uncertainties due to the scales. The uncertainty that is included for the data is obtained by adding all sources of experimental uncertainties in quadrature. In the lower insets, the ratio of the predictions NNLO+N$^{3}$LL-to-NNLO+NNLL and where applicable the predictions to the data for the $p_{T}$ are provided.}
\end{figure}

In this appendix, the predicted distributions for the differential cross section as a function of the forward Z boson $p_{T}$ up to 270 GeV are provided at 8, 13, and 14 TeV. The same computational setup, fiducial acceptance, uncertainty estimation procedure are used as discussed in the main body paper. The combined predictions are reported for the NNLO+NNLL and NNLO+N$^{3}$LL accuracies and compared to the available LHCb dimuon data~\cite{Aaij:2016mgv} at 13 TeV as well in figure.~\ref{fig:6}. The predictions are able to provide reasonable description of the $p_{T}$ distribution within uncertainties. The NNLO+N$^{3}$LL tends to predict lower (higher) $p_{T}$ distribution in the low-$p_{T}$ (intermediate- to high-$p_{T}$) region in comparison to the NNLO+NNLL prediction, which assures a more reliable description of the Z boson $p_{T}$ spectrum in the forward acceptance region.     

\acknowledgments
We wish to thank Marius Wiesemann, one of the authors of the MATRIX+RADISH, for providing valuable helps to set up the computational framework.

\end{document}